\documentclass[reqno]{amsart} \usepackage{amscd}
\usepackage{epsfig}
\newtheorem{theorem}{Theorem}[section]
\newtheorem{proposition}[theorem]{Proposition}
\newtheorem{lemma}[theorem]{Lemma}

\theoremstyle{definition}
\newtheorem{definition}[theorem]{Definition}

\theoremstyle{remark} \newtheorem{remark}[theorem]{Remark}
\numberwithin{equation}{section}
\newcommand{\be}{\begin{eqnarray}} 
\newcommand{\ee}{\end{eqnarray}}
\newcommand{\field}[1]{\ensuremath{\mathbb{#1}}}
\newcommand{\CC}{\field{C}}

\newcommand{\HH}{\field{H}}
\newcommand{\PP}{\field{P}}
\newcommand{\RR}{\field{R}}
\newcommand{\TT}{\field{T}}
\newcommand{\ZZ}{\field{Z}}
\newcommand{\MM}{\field{M}}
\newcommand{\NN}{\field{N}}
\newcommand{\LL}{L^+}        
\DeclareMathOperator{\Op}{\hat{\mathcal O}}

\DeclareMathOperator{\Tr}{Tr}

\DeclareMathOperator{\SL}{SL}

\DeclareMathOperator{\SO}{SO}
\DeclareMathOperator{\SU}{SU}
\DeclareMathOperator{\so}{\mathfrak so}
\begin{document}
%
%
\title[Twistors and Holography]{Twistors, CFT and Holography}
\author{Kirill Krasnov} \address{Albert Einstein Institute\\ Golm/Potsdam, 14476\\ Germany} 
\email{krasnov@aei.mpg.de}
\begin{abstract}
According to one of many equivalent definitions of twistors a (null) twistor is a null geodesic in Minkowski spacetime. Null geodesics can
intersect at points (events). The idea of Penrose was to think of a spacetime point as a derived concept: points are obtained
by considering the incidence of twistors. One needs two twistors to obtain a point. Twistor is thus 
a ``square root'' of a point. 
In the present paper we entertain the idea of quantizing the space of twistors. Twistors, and thus also spacetime points 
become operators acting in a certain Hilbert space. The algebra of functions on spacetime becomes an operator
algebra. We are therefore led to the realm of non-commutative geometry. This non-commutative geometry
turns out to be related to conformal field theory and holography. Our construction sheds an interesting new
light on bulk/boundary dualities. 
\end{abstract}
\maketitle
\section{Introduction}
\label{sec:intr}

This paper resulted from author's attempt to understand what twistors are. We hope that our interpretation 
will be of interest to others. Useful sources on twistors are:  
the book by Huggett and Tod~\cite{book}, and the two volume set by
Penrose and Rindler~\cite{PR}. There are many equivalent definitions of a twistor. Both sources just cited emphasize
the relation to spinors and to the geometry of Minkowski spacetime. It is not too often emphasized, however,  
that an analog of twistors exists in any spacetime dimensions. Moreover, there is an analogous concept even
for the spaces $\RR^n$ with the usual flat positive definite metric. These analogs are well known to mathematicians.
When viewed in this generality it comes to ones attention that twistor space is always a symplectic manifold. This
fact suggests to try to quantize twistors. Since twistors are related to spacetime (or space) points, such a quantization
promotes points into operators. This brings non-commutative geometry into play. What makes such a quantization
interesting is that the resulting non-commutative geometry turns out to be related to conformal field theory and
holography! 

Much of the AdS/CFT correspondence is just the representation theory of the conformal group. Twistors are
also deeply rooted in the conformal group. From this perspective it is not very surprising that there
is a relation between holography, conformal field theory and twistors. We hope, however, that exhibiting this relation will
prove useful for a better understanding of holography. We also hope that our interpretation 
will attract to twistors more attention, which they certainly deserve!

It would be hard to summarize the results of this paper in the introduction; it would amount to repeating much of the
paper. Let us however at least state the way that the non-commutative geometry we obtain gets related to
CFT. As we have said, quantizing the twistor space we make the geometry of the original space (spacetime)
non-commutative. Functions on space become operators acting in a certain Hilbert space. We consider, in particular,
functions of a fixed conformal weight $\Delta$. Let ${\mathcal O}_\Delta(p)$ be such a function. Let $\hat{\mathcal O}_\Delta$
be the corresponding operator. We show that the trace of the operator product:
\begin{equation*}
\Tr\left(\Op_{\Delta_1} \dots \Op_{\Delta_k}\right) 
\end{equation*}
satisfies all the properties of a conformal field theory correlation function. Our quantization procedure gives a way
to calculate all such quantities, and thus defines a particular conformal field theory. By ``defines'' here we mean that
one knows all of CFT's correlation functions; we don't know whether a Lagrangian formulation
exists. Finally, we speculate that a holographic bulk theory exists that dual to the CFT obtained.
We give a prescription for how one can find this dual theory.

Before we start, we would like to emphasize one important point. What is called twistor in our paper is a null twistor according to
more standard definitions. ``Usual'' twistors form a linear space, while our
(null) twistors do not. However, they form an orbit in the Lie
algebra of the conformal group. This orbit can be naturally quantized, and this eventually leads to a CFT
interpretation. One could have also attempted to quantize the space of ``usual'' twistors, making use of the
fact that it is a linear space. Such quantization have been considered, e.g.~\cite{Penrose} and references therein.
The arising picture seems very different from ours. Thus, one of the main novelties of the present paper is
the suggestion to treat the space of {\it null} twistors as the central object. Some advocates of twistor
theory may consider this as being against the whole philosophy of twistors. However, as we shall attempt to
demonstrate, this idea leads to a very natural construction.

Twistors exist in any dimension. However, only in dimensions $n=2,4,8$ is the twistor space a complex manifold. This
is due to the fact that only in this dimensions there is a ``coincidence'' and the real conformal group
$\SO(1,n+1)$ is locally isomorphic to a complex group. Twistor constructions in 4d use the fact that $\TT$ is a complex
space in a crucial way. Our construction, being possible in any dimension, only uses the real structure on $\TT$.
Again, this may be viewed as going against the twistor philosophy. Still, we hope that the naturalness of our construction
justifies our approach. 

Let us however keep this introduction short
and jump right to business. The paper is organized as follows. We start by reviewing various definitions of twistors.
The main aim of section~\ref{sec:twistors} is to understand the relation of twistors to spacetime (space) geometry
and to describe the space of twistors as a symplectic manifold.
We quantize this symplectic manifold in section~\ref{sec:quant}. Section~\ref{sec:rep} considers the question of the
decomposition of the tensor product of the twistor representation with its dual, and introduces an important graphical
calculus. The CFT and holographic interpretations are presented in
sections~\ref{sec:cft} and~\ref{sec:hol} correspondingly. We conclude with a discussion.

\section{Twistors}
\label{sec:twistors}

\subsection{}
The idea of the twistor programme of Penrose~\cite{Penrose} is that twistors should be thought as more fundamental than 
spacetime points. One needs two twistors to define an event. Twistors are thus ``square roots'' of spacetime points.
Via the help of the Penrose transform solutions of the zero mass wave equation can be coded into holomorphic functions of
an appropriate degree of homogeneity on the twistor space $\TT$, see e.g.~\cite{book}. Similarly, self-dual solutions of 
Yang-Mills equations (instantons) can be coded into holomorphic vector bundles over $\TT$, see e.g.~\cite{Atiyah}. 
Self-dual solutions of Einstein equations (Penrose's non-linear graviton) can be coded into complex structures on $\TT$, 
see~\cite{book}. Thus, the twistor formulation is certainly handy in describing the spaces of solutions of various 
field equations. The idea of Penrose was to base on twistors also the quantum fields, and, in particular, quantum gravity.
This programme has led to some important achievements; however some open problems remain. It is not our aim to 
review here the status of the twistor programme; see the available rich literature on the subject. In this paper, instead
of quantizing fields, we shall consider a quantization of the twistor space itself. 
At this stage the only justification for doing this is that it is possible. However, at the end we will arrive at
a certain CFT interpretation. In this section, as a pre-requisite for quantization, we shall describe how 
the twistor space is a symplectic manifold. 

The notion of twistor allows generalizations to other than 4-dimensional spacetimes. We need such a generalization
if we are to relate twistors to holography, for holography is not limited to a particular number of dimensions.
Moreover, analogs of twistors can be defined for the usual $\RR^n$. The AdS/CFT dualities are most cleanly set up 
in the case of Euclidean signature. Thus, because our final goal is to provide a holographic interpretation, we will
mostly work with these Euclidean analogs of twistors. This will be quite sufficient to illustrate the main idea, and will also 
allow us to greatly simplify the presentation. Everything we say about Euclidean twistors generalizes almost
immediately to the Lorentzian case. Because of such a more general viewpoint the definition of twistors that we give
is different from the usual definitions. In the case of 4 dimensions Lorentzian signature our definition is more or less
equivalent to the standard ones~\cite{book}.

\subsection{} Here we give a definition of twistors that is most useful for what follows. Our definition uses 
bivectors and Plucker relations. We explain the relation to
geometry, and in particular explain how points arise from the incidence of twistors. In the next subsection we explain how
our definition is related to the usual ones. 

\subsubsection{}
To define twistors we first need to introduce the conformal compactification of $\RR^n$. This is a standard construction.
Compare, e.g. with the discussion in Section 1.2 of~\cite{PR}. Consider Minkowski space
$\MM^{1,n+1}$. Let $\{x^0,x^1,\ldots,x^{n+1}\}$ be coordinates on $\MM^{1,n+1}$ so that the metric is
${\rm diag}(1,-1,\ldots,-1)$. Consider the positive light cone:
\be
{\LL}=\{x\in \MM^{1,n+1}: (x^0)^2-(x^1)^2-\ldots-(x^{n+1})^2=0, x^0>0 \}.
\ee
Conformal compactification of $\RR^n$ is defined to be the projective light cone $\PP{\LL}$, or the space of generators
of $\LL$. Equivalently, the conformal compactification is the section of $\LL$ by any spacelike hypersurface
that does not pass through the origin. It is obvious that topologically the compactification is $S^n$. The original space $\RR^n$
is also a section of $\LL$, but in this case by a null plane $x^0+x^{n+1}=1$. The embedding of $\RR^n$ into
$\MM^{1,n+1}$ is given explicitly by:
\be
x^a \to \left(\frac{1}{2} (1+(x,x)), x^a, \frac{1}{2}(1-(x,x)) \right).
\ee
Here $(x,x)=x_a x^a$. This embedding is an isometry. The generator $x^0+x^{n+1}=0, x^a=0$ of $\LL$ is a point at infinity 
that is added to $\RR^n$ to compactify it to $S^n$.

\subsubsection{}
We define twistor to be a pair:
\begin{definition} \label{def:pair} 
twistor = (point $p$ of $S^n$, vector in the tangent space at $p$).
\end{definition} 
\begin{figure}
\centering 
\epsfig{figure=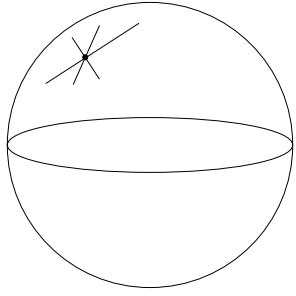, height=1.5in}
\caption{Twistor is a point $p$ on the sphere $S^n$ together with a tangent vector at $p$. Projective twistor is a point together
with a direction in the tangent space. Incident projective twistors correspond to directions through a fixed point. All such
twistors lie on a line $L_p$ in the projective space $\PP\TT$.}
\label{fig:sphere}
\end{figure}
%
A more practical, but
equivalent definition is as follows. Here indices take values from $0$ to $n+1$ and the metric of $\MM^{1,n+1}$ 
is used to lower and raise them.
\begin{definition}
Bivector $B_{ij}$ is an anti-symmetric tensor $B_{ij}=-B_{ji}$.
\end{definition}
\begin{definition}
Bivector is said to be simple if it is a skew product of two vectors $B_{ij}=2 u_{[i}v_{j]}$.
\end{definition}
Simple bivectors are in correspondence with 2-planes $\mu u_i +\nu v_i, \mu,\nu\in\RR$. We have the following well-known fact:
\begin{lemma}
The necessary and sufficient condition for a bivector to be simple is 
\be\label{simplicity}
B_{[ij} B_{kl]} = 0.
\ee 
\begin{proof} Let us first note that this condition is equivalent to:
\be\label{simplicity'}
B_{[ij} B_{k]l} = 0.
\ee
The fact that this is a necessary condition can be verified directly. Let us show that if this condition is satisfied then
the bivector factorizes. If $B_{ij}$ is not identically zero, we can find two vectors $a^i, b^i$ such that $B_{ij} a^i b^j=1$. Define
$u_i=B_{ij}a^j, v_i=B_{ij} b^j$. We have:
\begin{equation*}
u_i v_j-v_i u_j = (B_{ik} B_{jl} - B_{il} B_{jk}) a^k b^l = B_{ij} B_{kl} a^k b^l = B_{ij}
\end{equation*}
Here to pass from the second to third expression we have used the simplicity condition.
\end{proof}
\end{lemma}
We are now ready to define twistors:
\begin{definition}
Twistor is a simple null bivector, in other words a simple bivector that satisfies:
\be\label{null}
B_{ij} B^{ij} =0.
\ee 
Projective twistor is a simple null bivector up to scale.
\end{definition}

\begin{remark} Let us note that the twistor space $\TT$ as given by our definition is not a linear space: 
the sum of two null simple bivectors is not always a null simple bivector. As we shall see, this does not cause any problems
for our construction. We note that our twistors is what would
be usually called null twistors.
\end{remark}

We see that the twistor space $\TT$ is a quadric~\eqref{simplicity},~\eqref{null} in the space of bivectors
${\mathcal B}\sim \RR^{(n+2)(n+1)/2}$. We note that $\mathcal B$ coincides with the Lie algebra of the
conformal group ${\mathcal G}=\so(1,n+1)$. The space of simple bivectors in $\mathcal B$ 
consists of a set of orbits under the adjoint action of the group. Different orbits are distinguished by the value of $B_{ij}B^{ij}$. 
The twistor space $\TT$ is the null orbit $B_{ij}B^{ij}=0$. To determine the stabilizer, let us take a particular
simple null bivector, the one with $u=(1,0,\ldots,0,1), v=(0,0,\ldots,1,0)$. By looking at the corresponding 
matrix for $B$ it is easy to see that the stabilizer is $\SO(1,n-1)\times\SO(2)$. Thus, the twistor space can
be described as the following homogeneous group manifold:
\be\label{orbit}
\TT = \SO(1,n+1)/\SO(1,n-1)\times\SO(2), \qquad \dim\TT=2n.
\ee
The fact that the twistor space $\TT$ is an orbit will serve as a starting point for quantization.

\begin{remark} When $n=2$ we have the usual Minkowski space $\RR^{1,3}$. In this case the compactified space is $S^2$.
The group $\SO(1,3)\sim \SL(2,\CC)$ is the conformal group. Twistors are the usual spinors in this case.
Spinors can be described geometrically as a null direction (flag pole) together with a vector tangent to $S^2$,
see Section 1.4 of \cite{PR}. This information can be encoded in a null 2-plane that passes through the origin. This is why 2-planes 
are relevant. 
\end{remark}

\subsubsection{} Let us now explain the relation between simple null bivectors and pairs~\ref{def:pair}. The following discussion
is similar to that of Section 1.4 of \cite{PR}. We first need a notion of null planes. Any 2-plane $\mu u_i + \nu v_i, 
\mu, \nu\in \RR, u_i, v_i \in \MM^{1,n+1}$ can have at most two null directions. These are the real solutions of the
quadratic equation:
\begin{equation*}
\mu^2 (u,u) + \nu^2 (v,v) + 2\mu\nu (u,v)=0.
\end{equation*}
When these null directions coincide the plane is called null. Let us choose $u$ to be this null direction. Then there is no
other null direction only if $(u,v)=0$, i.e. any other vector in the plane is orthogonal to $u$. We have the following:
\begin{lemma} The necessary and sufficient condition for a simple bivector $B_{ij}=2u_{[i}v_{j]}$ to represent a null 
plane is that this bivector is null.
\begin{proof} The proof follows obviously from:
\be\label{B-squared}
B_{ij} B^{ij} = 2 (u,u) (v,v) - 2 (u,v)^2.
\ee
\end{proof}
\end{lemma}
Thus null simple bivectors correspond to null 2-planes through the origin. We could have therefore defined twistors as 
null 2-planes through the origin. The unique null direction of this plane determines
a point $p$ on $\PP\LL$, that is a point of the conformal compactification of $\RR^n$. 
The other orthogonal direction $v$ is a spacelike vector
which determines a tangent vector at $p$. Thus, null 2-planes (twistors) are indeed just pairs of definition~\ref{def:pair}.

\begin{remark}\label{2-planes}
Let us introduce, for future reference, a notion of a timelike 2-plane. This is a plane with two non-coinciding
null directions. Let $B$ the corresponding simple bivector. We shall also call it timelike. 
Let us choose $u, v$ to be the null directions. Then, as we see from~\eqref{B-squared},
for $B$ timelike $B_{ij}B^{ij}<0$. Spacelike 2-planes are those without null directions. The corresponding bivectors
$B: B_{ij}B^{ij}>0$.
\end{remark}

\subsubsection{} 
\begin{figure}
\centering 
\epsfig{figure=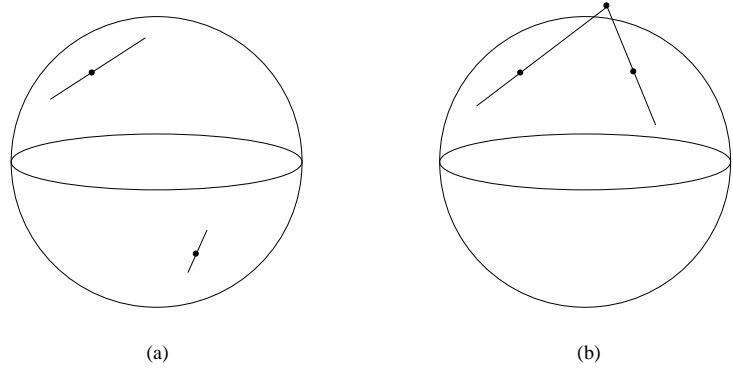, height=1.9in}
\caption{Two twistors do not have to intersect (a), or may intersect at an ``imaginary'' point (b).}
\label{fig:incidence}
\end{figure}
%
Now we would like to discuss the incidence of twistors. This is necessary if we want to understand how
twistors define spatial points. As Fig.~\ref{fig:incidence} shows, the 2-planes that represent twistors
do not have to intersect along a line. If they do intersect along a line, this line is generically spacelike, not
null. Thus, we need conditions to guarantee real incidence. These conditions are given by the following two lemmas:
\begin{lemma} Two 2-planes corresponding to simple bivectors $B, B'$ share a common direction (intersect along a line)
iff 
\be\label{inters}
B_{[ij}{B'}_{kl]}=0.
\ee
\begin{proof} 
First, given two simple bivectors $B_{ij}=u_{[i} v_{j]}, B'_{ij}=v_{[i} w_{j]}$, any linear combination of $B, B'$ is a
simple bivector. In view of~\eqref{simplicity} this is only possible if~\eqref{inters} holds. To prove that~\eqref{inters}
is a sufficient condition we construct the common direction explicitly. Consider $B_{i[j} B'_{kl]}$. If the two bivectors
are proportional, then this expression vanishes identically. Otherwise it is possible to find vectors $a^i, b^i, c^i$ such
that:
\begin{equation*}
v_i = B_{i[j} B'_{kl]} a^j b^k c^l
\end{equation*}
is different from zero. It is easy to see that such $v_i$ is the common factor of $B$. Indeed, we have
\begin{equation*}
B_{[ij} v_{k]} = B_{[ij} B_{k][l} B'_{mn]} a^l b^m c^n = 0.
\end{equation*}
In view of the Plucker relation~\eqref{inters} 
\begin{equation*}
B_{i[j} B'_{kl]}+ B'_{i[j} B_{kl]}=0
\end{equation*}
so the vector that divides $B'$ coincides with the one that divides $B$.
\end{proof}
\end{lemma}
\begin{lemma} Two null 2-planes corresponding to null simple bivectors $B, B'$ intersect along a null line iff
in addition to~\eqref{inters} the following relation holds:
\be\label{null-inters}
B_{ij} {B'}^{ij} = 0.
\ee
\begin{proof} We have:
\begin{equation*}
B_{ij} {B'}^{ij} = 2(u,u')(v,v') - 2(u,v')(u',v).
\end{equation*}
Thus, if $u=u'$ is the common null direction then~\eqref{null-inters} holds. To prove the converse assume that
the planes intersect along some direction that is not null. Choose this common direction to be $v=v'$. Then~\eqref{null-inters}
gives $(u,u')(v,v')=0$. Because we have assumed that $u\not= u'$ this means that $(v,v')=0$, which is a contradiction.
Thus, $u=u'$ must be the common direction.
\end{proof}
\end{lemma}

Recall now that a null line being a generator of $\LL$ is a point of the compactified $\RR^n$.
This is the way to represent points in the twistor language: a spatial point is represented by a pair of twistors (null simple 
bivectors) that intersect along a null line. However, if two simple null bivectors intersect along a null line, then any 
linear combinations of them will have the same property. Thus, points $p$ actually correspond to two dimensional
surfaces $\mu B + \nu B'$ in the twistor space. Different twistors on this surface correspond to different 
tangent vectors at $p$. It is often convenient to pass to the projective twistor space. Points are
represented by lines $L_p$ in the projective twistor space $\PP\TT$. Different projective twistors on
this line are different tangential directions at $p$, see Fig.~\ref{fig:sphere}.

\subsection{} The definition of twistors that we gave is sufficient for our purposes. Moreover, we have already seen
that the twistor space $\TT$, being an orbit in the Lie algebra, is naturally a symplectic manifold. Thus, we could
proceed to quantization. Let us however first explain why our definition is equivalent to the standard ones. 

The usual definition of twistors in Minkowski space $\MM^{1,3}$, see \cite{book}, relies on the fact that the conformal group 
$\SO(2,4)$ is (locally) the same as $\SU(2,2)$. Twistors can then be identified with vectors in the spinorial representation of this 
group. It has complex dimension 4, so twistors are pairs of spinors $Z^\alpha=(\omega^A,\pi_{A'})$. Null twistors are defined as
those for which $Z^\alpha Z_\alpha=0$, where one lowers and raises indices with a metric of signature $(+,+,-,-)$. 
Projective null twistors correspond to null geodesics in the compactified $\MM^{1,3}$. Two null geodesics corresponding
to $X,Y$ intersect iff $X^\alpha \bar{Y}_\alpha=0$. If two twistors intersect then any their linear combinations do. Thus,
there is a line $L_p$ in the projective null twistor space $\PP\NN\TT$ such that twistors on this line correspond to null geodesics
intersecting at $p$. In other words, $L_p$ represents the null cone of $p$. Completely the same structure was coded into
our definition, except that we worked with Euclidean twistors. Note, however, that our definition works in any dimension
because it does not use the coincidence $\SO(2,4)\sim\SU(2,2)$.

The advantage of the usual ``spinorial'' definition is that it is rather easy to obtain the dual description. Namely,
lines in $\PP\TT$ can be represented
by the corresponding simple bivectors $P^{\alpha\beta} = 2X^{[\alpha} Y^{\beta]}$. A bivector $P^{\alpha\beta}$ is simple
iff 
\be\label{simple-twistor}
\epsilon_{\alpha\beta\gamma\delta} P^{\alpha\beta} P^{\gamma\delta}=0. 
\ee
Thus, the space of lines on $\PP\TT$ becomes identified with the quadric~\eqref{simple-twistor} in the space of projective
bivectors. It is moreover easy to show that all points on a line corresponding to simple $P^{\alpha\beta}$ are null iff
\be\label{null-twistor}
\bar{P}_{\alpha\beta}=\frac{1}{2} \epsilon_{\alpha\beta\gamma\delta} P^{\gamma\delta}.
\ee
The space of bivectors satisfying~\eqref{null-twistor} is the real projective space $\RR\PP^5$. Simple bivectors~\eqref{simple-twistor}
form a quadric in $\RR\PP^5$. Since such bivectors are exactly the lines $L_p$ in $\PP\NN\TT$ that correspond to points we
get that the compactified Minkowski space is represented as the quadric~\eqref{simple-twistor} in $\RR\PP^5$. This is the
so-called Klein correspondence. It is of central importance for the Penrose transform. 

The definition that we used based on bivectors also gives 
the dual description: points $p$ of the compactified $\RR^n$ correspond to lines $L_p$ in $\PP\TT$. It is possible to
give a direct description of the space of lines $L_p$ using bivectors. As the result one gets a quadric in $\RR\PP^{n+1}$, 
which is the compactified $\RR^n$. We will not need this explicitly in the present paper.

\section{Quantization}
\label{sec:quant}

\subsection{}
The way in which we described the twistor space makes it clear that $\TT$ is a symplectic manifold. Indeed, it is
an orbit~\eqref{orbit} in the Lie algebra, and so the usual Kirillov-Kostant symplectic form exists
on $\TT$. It is the unique symplectic form on $\TT$ that is invariant under the conformal group 
transformations. It is this symplectic form that we will use to quantize $\TT$.

\begin{remark} In the usual ``spinorial'' description of twistors another symplectic form appears very naturally,
see, e.g.,~\cite{Penrose}. The corresponding Poisson brackets are given by:
\begin{equation*}
\{Z^\alpha,Z^\beta\}=0,\qquad \{\bar{Z}_\alpha,\bar{Z}_\beta\}=0, \qquad \{Z^\alpha,\bar{Z}_\beta\}=\delta^\alpha_\beta.
\end{equation*}
This means that the complex conjugate twistor is also the canonically conjugate. This structure plays an
important role in the Penrose transform. We do not know whether this symplectic structure on the full twistor
space has anything to do with the orbit symplectic structure that only arises after one restricts one's attention
to the null twistors. In this paper we consider a quantization of the space of null twistors only.
\end{remark}
\begin{remark} The ``usual'' twistor space $\PP\TT$ of projective Lorentzian 4-dimensional twistors is a complex projective
plane $\CC\PP^3$. Non-commutative deformations of projective planes have been studied. 
The work~\cite{Kapustin} relates the non-commutative
geometry arising from such a deformation to the ``usual'' one
with the Moyal non-commutative product of functions. The non-commutative
geometry that arises from our construction is different. 
\end{remark}

\subsection{} What we have to quantize is the space of simple null elements in the Lie algebra $\so(1,n+1)$. Such elements become
operators in some Hilbert space $\mathcal H$. This Hilbert space gives some representation of the conformal group. It is natural to ask
which irreducible representations appear in the decomposition of $\mathcal H$. Another way to pose the problem is: find all irreducible 
representations $V^\rho$ of the conformal group such that the operators $X_{[ij}X_{kl]}, X_{ij}X^{ij}, X_{ij}\in\so(1,n+1)$ are
zero operators in $V^\rho$. In fact, since the space of null simple bivectors is a single orbit, the method of orbits suggests
that there is just a single representations $V^\rho$ with the desired properties. We would like to find this representation.

A similar problem has been considered in the framework of spin foam models of quantum gravity. The idea there was to quantize a geometric
simplex in $\RR^n$ by quantizing its 2-faces. This idea was first proposed by Barbieri~\cite{Barbieri} in 3-dimensions, and then applied
to 4-simplexes by Barrett and Crane~\cite{BC} and simultaneously by Baez~\cite{Baez}. The representation theoretic meaning of these
constructions was clarified in~\cite{FKP}. It is this last reference that is most relevant for our purposes. The relevant construction
is as follows. Consider 2-planes in $\RR^n$. They can be described as simple bivectors. Bivectors can be identified with elements of
the Lie algebra $\so(n)$. The problem is thus to find irreducible representations of $\SO(n)$ that solve the simplicity condition.
It was shown that these are exactly the so-called simple, or representations that can be realized as spherical harmonics 
on $S^{n-1}$. Analogous representations will appear in our case. 

\subsection{}
Let us remind the reader how some of the representations of $\SO(1,n+1)$ are constructed. Our reference here is the book by
Vilenkin and Klimyk~\cite{VK}, Vol. 2. The representations we are about to describe are of the most degenerate series when only
one of the Casimirs is non-zero. They are also the most studied ones. Fortunately,
they are exactly the representations we need. Let us denote by ${\mathfrak B}^\sigma$ the space of functions on the positive
light cone $\LL$ homogeneous of degree $\sigma$. In other words, we consider functions such that $f(t u)=t^\sigma f(u), u\in\LL, t>0$.
This space is obviously invariant with respect to the shift operators $(T(g) f)(u) = f(g^{-1} u)$. We denote the corresponding
representation by $V^\sigma$.

The same representations can be realized in the space of homogeneous {\it harmonic} functions inside (outside) of the
light cone. The relation between the two realizations is as follows: a homogeneous function on $\LL$ can be used as
the boundary data for the equation $\partial^i \partial_i f = 0$ inside $\LL$. Solving this equation one gets a homogeneous harmonic
function inside $\LL$. In this second realization of the representations, the Lie algebra elements are represented
as:
\be\label{X}
X_{ij} = x_i \frac{\partial}{\partial x_j} - x_j \frac{\partial}{\partial x_i}.
\ee
It is now obvious that all representations $V^\sigma$ satisfy the simplicity condition. Indeed:
\be
X_{[ij} X_{kl]} f = 4 x_{[i} \partial_{j} x_k \partial_{l]} f = 4 x_{[i} \delta_{jk} \partial_{l]} f + 4 x_{[i} x_j \partial_k \partial_{l]} f = 0.
\ee
Let us now consider the null condition. Consider the quadratic Casimir:
\be
C= \frac{1}{2} X_{ij} X^{ij} = (x,x) \partial^i \partial_i - n x^i\partial_i - x^i x^j \partial_i \partial_j.
\ee
When acting on $f\in V^\sigma, x^i \partial_i f = \sigma f$. Also, because the functions are harmonic $\partial^i \partial_i f=0$,
so we have:
\be\label{casimir}
X_{ij} X^{ij} f = - 2\sigma (\sigma + n) f.
\ee
We need a representation in which $C$ is the zero operator. There are two values of $\sigma$ that give that:
$\sigma=0$ and $\sigma=-n$. As we shall see, the two representations are duals (with respect to a bilinear form) of each other, so it
is enough to consider one of them. We take it to be $V^0$. In words, this is the representation in the space of 
functions on $\LL$ of degree of homogeneity zero. 

Let us give
another, equivalent description. Homogeneous functions inside the light cone are determined by their value on any spatial
hypersurface. In particular, one can take the hypersurface $(u,u)=1$, which is nothing but the hyperbolic space $\HH^{n+1}$.
Homogeneous harmonic functions inside $\LL$ are therefore the same as eigenfunctions of the Laplacian on $\HH^{n+1}$. On the
space $V^\sigma$ the equation $\partial^i \partial_i f = 0$ reduces to:
\be\label{laplace}
\Delta_\HH f = \sigma(\sigma+n) f.
\ee
Thus, the representation space $V^0$ can also be described as the space of massless fields in $\HH^{n+1}$:
\be
V^0=\{f: \Delta_\HH f = 0\}.
\ee
The holographic interpretation that we shall describe arises from this realization.

\begin{remark} We note that the remaining simple representations, namely those with $\sigma=l, l\in\ZZ_+$ (the quantization of $\sigma$ is
necessary for unitarity, see below), describe the timelike 2-planes. Indeed, first, they 
all satisfy the simplicity condition, so they describe 2-planes. Second, as it
is clear from~\eqref{casimir}, the Casimir $X_{ij}X^{ij}$ takes negative values on $V^l=V^\sigma, \sigma=l, l\in\ZZ_+$.
Because the Casimir represents $B_{ij}B^{ij}$, recalling the remark~\ref{2-planes}, 
we see that the representations of the discrete series correspond to timelike 2-planes.
We also note that, as is clear from~\eqref{laplace}, representations of the discrete series can be realized in the
space of massive fields on $\HH^{n+1}$:
\be\label{wave}
\Delta_{\HH} f_l = m_l^2 f_l, \qquad m_l^2 = l(l+n). 
\ee
All representations $V^l$ are lowest weight, so, using the AdS/CFT language, their energy is bounded from below. Being described
by fields of positive $m^2$ they are exactly the representations that have interpretation in AdS/CFT correspondence. We shall
return to this point below, when we talk about the holographic interpretation.
\end{remark}

\subsection{} Let us review some more facts about the representation $V^0$. It lies on the intersection of the so-called
discrete and supplementary series of the unitary irreducible representations of $\SO(1,n+1)$. The former correspond to 
$\sigma = l, l\in\ZZ_+$ and the later are $-n < \sigma <0$, see~\cite{VK}, Vol. 2, so the point $\sigma=0$ is
indeed their intersection. However, strictly speaking, it belongs to neither one of the two series. It is the so-called
unipotent representation. For a general theory of unipotent representations see \cite{Vogan}. Another
remark is that the representations $V^l$ are not irreducible. There is an invariant subspace consisting of 
polynomials of degree $l$. Thus, strictly speaking, the representation one has to consider is the quotient representation,
see~\cite{VK}.

The inner product in $V^\sigma$ is easiest to describe in the realization in the space of homogeneous functions on $\LL$.
Let $\phi, \psi \in V^\sigma$ be two vectors. Their inner product is given by:
\be\label{ip*}
(\phi, \psi) = \int_{S^n} d\xi \int_{S^n} d\eta \,\, (\xi,\eta)^{-n-\sigma} \overline{\phi(\xi)} \psi(\eta).
\ee
Here $\xi, \eta$ are vectors on $\LL$. The expression under the integrand is invariant under re-scalings of $\xi, \eta$,
so it is a function on the two copies of the projective light cone. We note that the kernel that is used to write
an invariant expression is just the CFT 2-point function for operators of conformal dimension $\Delta=n+\sigma$. Such operators
are dual to functions $\psi_\sigma\in V^\sigma$ with respect to the pairing:
\be
\int_{S^n} d\xi \Op_{-n-\sigma}(\xi) \psi_\sigma(\xi).
\ee
Similarly, one can also describe the inner product \eqref{ip*} as a pairing between the functions from $V^\sigma$ 
and the functions from the dual space $V^{-n-\sigma}$. Indeed, consider a mapping:
\be\label{mapping-dual}
A: V^\sigma \to V^{-n-\sigma}, \qquad (A\psi)(p) =\int_{\RR^n} dq\, |p-q|^{-2(n+\sigma)} \psi(q).
\ee
Now, the inner product is just a pairing:
\be\label{ip}
(\phi, \psi) = \int_{\RR^n} dp\,\, \overline{\phi(p)}(A\psi)(p).
\ee
It is not at all obvious that the integral that is used to define the mapping \eqref{mapping-dual} makes sense. 
The kernel $|p-q|^{-2(n+\sigma)}$ is a generalized function. When $\sigma=l$, which corresponds to the
representations of the discrete series, the kernel has a simple pole. The residue at this pole is a 
differential operator. The polynomials of degree $l$ are
in the kernel of the mapping \eqref{mapping-dual}. The pairing \eqref{ip} reduces to an inner product on the quotient space. 
The mapping inverse to \eqref{mapping-dual} is realized as an integral operator. 

The case of $\sigma=0$, which is our twistor representation, requires even more care. The above description of the
inner product for the discrete series $\sigma=l$ does not extend to the case $\sigma=0$. To give an explicit realization of
this representation we need to describe a certain other realization of the discrete series. This is the twistor realization.
We will not need any details of this in the present paper, so we shall limit our presentation to a description of
the basic idea. The twistor realization of the discrete series is in terms of functions on the twistor space. This
realization is very well known for the case of $n=1$. The group in this case is $\SO(1,2)\sim\SL(2,\RR)$. The twistor
orbit \eqref{orbit} in this case is the unit disk. As is well known, the discrete series of representations of
$\SL(2,\RR)$ can be realized in the space of functions analytic inside the unit disk. The inner product is given by
an integral over the interior of the unit disk (it can also be represented as an integral over the boundary circle).
What is of interest for us is the representation $V^0$ in this case. Its explicit realization is in terms
of functions analytic inside the unit disk, with the inner product given by:
\be
(f,g) = \int d^2z \overline{f(z)} g(z).
\ee
For the representation $V^0$ there is no nontrivial kernel inside the integral. For more details on 
the twistor realizations one can consult \cite{Baston}. In what follows we will not need any of these
details. The purpose of the above discussion was simply to convince the reader that  
the twistor representation $V^0$ exists.

\subsection{} The representation $V^0$ satisfies the conditions that the operators $X_{[ij}X_{kl]}$ and $X_{ij}X^{ij}$, 
$X_{ij}\in\so(1,n+1)$ are
zero in $V^0$ and therefore gives the quantization of the twistor space.
Thus, $V^0$ is the Hilbert space in which operators corresponding to functions on $S^n$ will act. To construct such operators
explicitly we use the following standard construction. Let us consider the tensor product of representation $V^0$ with
its conjugate. Let us decompose this tensor product into irreducibles:
\be\label{dec}
V^0 \otimes \overline{V^0} = \oplus_\rho V^\rho.
\ee
Later we will show that the sum on the right hand side is indeed a sum, not a direct integral. Let us assume 
that all representations that appear on the right hand side of~\eqref{dec}
can be realized in spaces of functions (not necessarily scalar) on $S^n$. We will justify this assumption later. 
In view of~\eqref{dec} all such functions become operators acting in $V^0$. Explicitly,
given a function $f_\rho(p)$ that is a vector in some $V^\rho$, the corresponding operator can be described by its kernel:
\be\label{quantization-f}
(\hat{\mathcal O}^f \phi)(p) = \int_{S^n} dp' \,\, {\mathcal O}^f(p,p') \phi(p').
\ee
This kernel $\hat{\mathcal O}^f(p,p')$ depends on which function $f_\rho(p)$ it represents. One can introduce another
set of local operators operator $\hat{\mathcal O}_{\rho}(p)$, so that:
\be
\hat{\mathcal O}^f = \int_{S^n} dp f_\rho(p) \hat{\mathcal O}_{\rho}(p).
\ee
As we shall demonstrate below, the operators $\hat{\mathcal O}_{\rho}(p)$ have the interpretation of CFT operators of
conformal dimension $\Delta_\rho$. For example, for the representations of the discrete series, which, as we shall see,
are among those appearing on the right hand side of \eqref{dec}, the conformal dimension is $\Delta_l=n+l$.

The operators $\hat{\mathcal O}_{\rho}(p)$ are explicitly given by certain Clebsch-Gordan coefficients. Indeed, one can rewrite 
the kernel in \eqref{quantization-f} as:
\be\label{clebsch}
{\mathcal O}^f(p,p') = \int_{S^n} dq\,\, f_\rho(q) C_\rho(q, p, p').
\ee
Now the object $C_\rho(q, p, p')$ is the Clebsch-Gordan coefficient $C^{\rho \, 0 \, \bar{0}}_{f  \phi\, \psi}$ 
that arises from \eqref{dec}.
Indeed, the isomorphism~\eqref{dec} is such that given two functions $\phi\in V^0, \psi\in \overline{V^0}$ one obtains another function 
$f\in V^\rho$ according to:
\be
f_\rho(q) = \int_{S^n} dp \int_{S^n} dp' \,\, \phi(p) \psi(p') C_\rho(q, p, p').
\ee
Thus, $\hat{\mathcal O}_{\rho}(q)$ is indeed just the Clebsch-Gordan coefficient $C_\rho(q, p, p')$.

\begin{remark} Essentially the same quantization procedure is used to obtain the so-called fuzzy sphere. Let us briefly remind the
reader this construction. For more details consult e.g.~\cite{Fuzzy}. Let us take representation $V^{N/2}$ of dimension $N+1$ of
$\SU(2)$. One has:
\be
V^{N/2}\otimes \overline{V^{N/2}} = \oplus_{l=0}^{N} V^l.
\ee
Representations appearing on the right hand side can be realized in the space of functions on the sphere $S^2$. Vectors in each
$V^l$ are just the spherical harmonics $Y^{lm}$. The above isomorphism maps each $Y^{lm}$ into an operator acting in $V^{N/2}$.
Since $V^{N/2}$ is finite dimensional all such operators are $(N+1)\times (N+1)$ matrices $[Y^{lm}]_{ij}$. These matrices are just the
Clebsch-Gordan coefficients $C^{l (N/2) (N/2)}_{m \,\, i \,\, j}$. Thus, each spherical harmonic on $S^2$ with $l\leq N$ becomes an 
operator. The usual product of functions is replaced by the matrix product. The result of such quantization is a non-commutative
manifold known as the fuzzy sphere. In the limit $N\to\infty$ the fuzzy sphere becomes the usual commutative $S^2$. 
Our construction is completely analogous, except that the representations that we use
are infinite dimensional. Note also that in our construction the representation $V^0$ that plays the role analogous to $V^{N/2}$ 
is unique: there is no parameter that can be taken to infinity to make the space commutative. In this respect the non-commutative
geometry arising from our construction is ``rigid''. There is no way to take a limit and to remove the non-commutativity.
\end{remark}

Thus, all functions on $S^n$ that appear
on the right hand side of~\eqref{dec} become operators in the Hilbert space $V^0$. The usual commutative product of functions
gets replaced by the non-commutative operator product. This $\star$-product, being essentially a matrix product, is associative. 
We note that there is also a direct way to check associativity. It turns out to be equivalent to the so-called pentagon 
identity in the representation theory. The
proof of associativity is contained in~\cite{Fuzzy}. This reference deals with the fuzzy sphere geometry, but the proof presented is
completely general and applies without any changes to our current situation. We will not repeat it here. Thus, the non-commutative
geometry obtained is an ``honest'' one, with an associative $\star$-product.

\section{Calculus of representations}
\label{sec:rep}

We have quantized the twistor space and thus the compactified space $S^n$. Functions on $S^n$ got promoted into operators.
At this stage, however, this construction has nothing to do with either CFT or holography. To get
closer to a CFT interpretation we have to understand in more detail what class of representations appears on the
right hand side of~\eqref{dec}. In this section we shall also develop a certain graphical calculus; this
will prove instrumental when we discuss a CFT interpretation. 

\subsection{} We would like to start by demonstrating that the following assertion holds:
\begin{proposition} The representations that appear in the decomposition into 
irreducibles of the tensor product of the null representation with its conjugate are (i) the null representation itself 
together with all representations of the discrete series; (ii) tensor representations that will be defined below:
\be\label{dec'}
V^0\otimes\overline{V^0} = V^0 + \oplus_{l\in\ZZ_+} V^l + \oplus_{l,J} V^{l,J}.
\ee
\end{proposition}
This proposition has two parts. First we would like to prove that the first two terms on the right hand side appear.
After this, we will explain what are the representations denoted by $V^{l,J}$, and demonstrate that they appear.

\begin{figure}
\centering 
\epsfig{figure=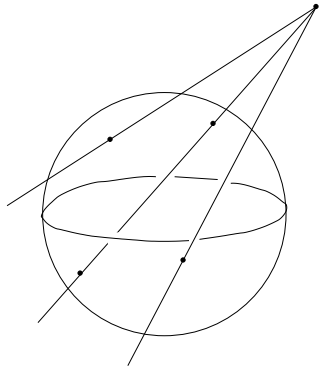, height=1.9in}
\caption{In the projective model 2-planes passing through the origin of the Minkowski space are represented by 
straight lines. Lines corresponding to null 2-planes just touch the sphere, 
those corresponding to timelike 2-planes intersect $S^n$ at two points. This figure shows lines corresponding to 3 planes:
2 null and one timelike, all sharing a spacelike direction. This direction in the projective model is represented by
a point outside of $S^n$ where lines intersect.}
\label{fig:projective}
\end{figure}

We start with the first part: the assertion that all simple representations appear on the right hand side of \eqref{dec'}.
We would like to give what can be called ``physicist's proof'' of this assertion. Our proof will be geometrical and based on 
simple bivectors in $\MM^{1,n+1}$. Indeed, we recall that the null representation $V^0$ describes 
simple null bivectors, that is null 2-planes through the origin. The representations of the discrete series describe 
timelike simple bivectors, that is timelike 2-planes through the origin. Let us work in the projective model obtained
by slicing $\MM^{1,n+1}$ by a spatial hypersurface $x^0=1$. In this projective description null 2-planes become lines
touching the sphere $S^n$, see Fig.~\ref{fig:sphere}. Timelike 2-planes become lines that pinch the sphere at two points,
see Fig.~\ref{fig:projective}.
%
If the intersection condition~\eqref{inters} is satisfied two 2-planes share a line, which is generically a spacelike one. This
line in the projective model is represented as a point outside of the sphere where lines corresponding to 2-planes intersect.

The projective model described is the basis for a well-known duality between points outside of the sphere and geodesic planes
in $\HH^{n+1}$. Although we are not going to use this duality in any essential way we describe it to make the geometry more clear. First we note
that the straight lines inside the sphere represent geodesics of $\HH^{n+1}$ in the projective model. The hyperbolic space
must be thought of as the interior of the sphere $S^n$. Thus, every point $p$ outside of the sphere gives rise to a whole set of
geodesics in $\HH^{n+1}$: these are the lines that intersect at $p$. All these geodesics are ultra-parallel in $\HH^{n+1}$.
Note now that there is a geodesic plane in $\HH^{n+1}$ that is orthogonal to all these geodesics. These geodesic plane intersects
the boundary sphere at a circle. This circle can be obtained by considering all lines through $p$ that touch the sphere.
The set of points where this happens is exactly the circle along which the geodesic plane intersects $S^n$. To summarize,
geodesic planes in the hyperbolic space are dual to points outside of the sphere.

To prove our assertion we take two null simple bivectors and consider their linear combination. Linear combinations of bivectors
correspond, at the level of representation theory, to tensor products of representations. This is familiar from the
quantum mechanics of angular momentum, when the sum of two angular momenta corresponds to the tensor product of two
representations. Generically bivectors do not share a line;
in the projective model this is the situation when the lines representing 2-planes do not meet. Such configurations of
null simple bivectors give rise to representations in the third term in~\eqref{dec'}. When null 2-planes are such that
they share a line, this line is generically spacelike. An arbitrary linear combination of the two bivectors is then a simple timelike
bivector. Simple timelike bivectors correspond to representations $V^l$ of the discrete series. This is why
they must appear in~\eqref{dec'}. Finally, the two null 2-planes may share a null direction. Then their linear combination
is also null. This explains the term $V^0$ on the right hand side of~\eqref{dec'}. This finishes the 
geometric proof. 

\subsection{}
A more direct way to prove~\eqref{dec'} is to consider the operator product expansion (OPE) of an operator $\Op_n$
of conformal dimension $\Delta=n$ with an operator $\Op_0$ of conformal dimension zero. Such OPE contains exactly the 
same information as the decomposition of the tensor product. The OPE is obtained using the Taylor expansion. 
One gets an infinite series in derivatives. The derivatives can be arranged into trace parts --these lead to
scalar operators of higher conformal dimension-- and traceless parts --these lead to tensor operators.
One can identify all these objects with the representations that appear on the right hand side of \eqref{dec'}. 
One finds that the non-simple representations on the right hand side of \eqref{dec'}
are tensor fields, or fields of non-zero twist.

Let us explain this statement in more detail. The representations that we denoted by $V^{l,J}$ 
are not simple in the sense that they do not
correspond to simple bivectors. They cannot be described with the help of scalar functions of some degree of homogeneity.
However, they can be described as appropriate tensors, or fields with non-zero twist. To see this, let us recall that:
\be
\HH^{n+1} = \SO(1,n+1)/\SO(n+1).
\ee
Representations of $\SO(1,n+1)$ can be induced from representations of $\SO(n+1)$. For example, there is a
representation of $\SO(1,n+1)$ in the space of symmetric traceless tensors fields of rank $J$ on $\LL$,
homogeneous of some degree $l$. Such tensor fields can be used to realize the remaining representations $V^{l,J}$ in~\eqref{dec'}. 
One can treat these tensor fields quite similarly to how scalars were treated. They get represented by operators, 
to know which one has to compute the Clebsch-Gordan
coefficients. For $\SO(1,5)$ The Clebsch-Gordan coefficients for some low twist fields were computed in the appendix of 
paper~\cite{Gleb}.  We note that the appearance of non-zero twist operators is usual in any example of AdS/CFT duality.

\begin{remark} It is illustrative to compare the decomposition~\eqref{dec'} with a similar decomposition for the so-called
singletons, see, e.g., \cite{Sing}. Singleton representation (originally discovered by Dirac for $\SO(2,3)$) can be
thought of as a ``square root'' of the massless representation. The tensor product of the singleton representation with itself
contains all massless representations of the conformal group. In contrast, our basic representation is massless, and the
tensor product contains all massive fields. The singleton representation does not seem to be relevant for our construction.
\end{remark}

\subsection{} Now we would like to develop a graphical calculus that will be used in an essential way in the next
section. We denote the identity operator in some representation space $V^\rho$ by a line. The operators
$\hat{\mathcal O}_{\rho}$, given by the Clebsch-Gordan coefficients 
$C^{\rho \, 0 \, \bar{0}}$, are represented by a 3-valent vertex: 
\be
\hat{\mathcal O}_{\rho} = C^{\rho \, 0 \, \bar{0}} = \lower0.4in\hbox{\epsfig{figure=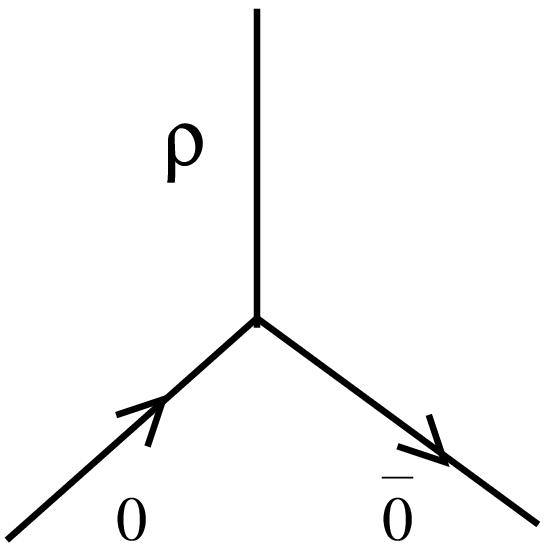,height=0.8in}}
\ee
The Clebsch-Gordan coefficients satisfy an orthogonality relation. Graphically this is represented as:
\be\label{3j-ort}
\lower0.35in\hbox{\epsfig{figure=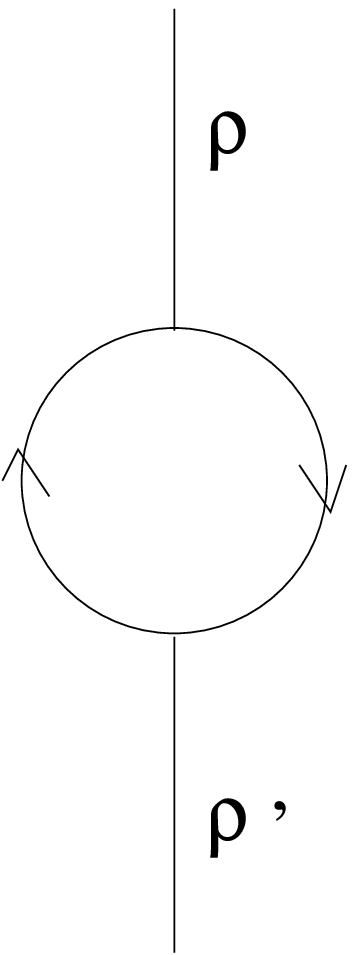,height=0.8in}}\,\,\, =\delta_{\overline{\rho} \rho'}  \,\,\, 
\lower0.25in\hbox{\epsfig{figure=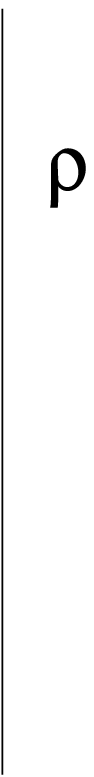,height=0.6in}}
\ee
The meaning of the quantity on the left hand side is that one multiplies two Clebsch-Gordan coefficients, and sums over
a basis of vectors in the spaces $V^0, \overline{V^0}$. The orthogonality relation \eqref{3j-ort} fixes the
normalization of the Clebsch-Gordan coefficients.

\section{CFT interpretation}
\label{sec:cft}

\subsection{} Having obtained the above preliminary results, we are ready to give a CFT interpretation. As we have
seen in section \ref{sec:quant}, a function of some degree of homogeneity $l$ becomes an operator acting in the Hilbert space $V^0$. 
The result of the multiplication of two such operators is obtained, in the graphical notation, by drawing next to
each other the two 3-valent vertexes representing the Clebsch-Gordans, and connecting two of the legs so that the arrows match. 
Thus, the trace of a product of a number of operators is equal to the following diagram:
\be\label{product-pr}
\Tr{\left(\Op_{\rho_1} \Op_{\rho_2} \Op_{\rho_3} \ldots \Op_{\rho_k}\right)} = 
\lower0.55in\hbox{\epsfig{figure=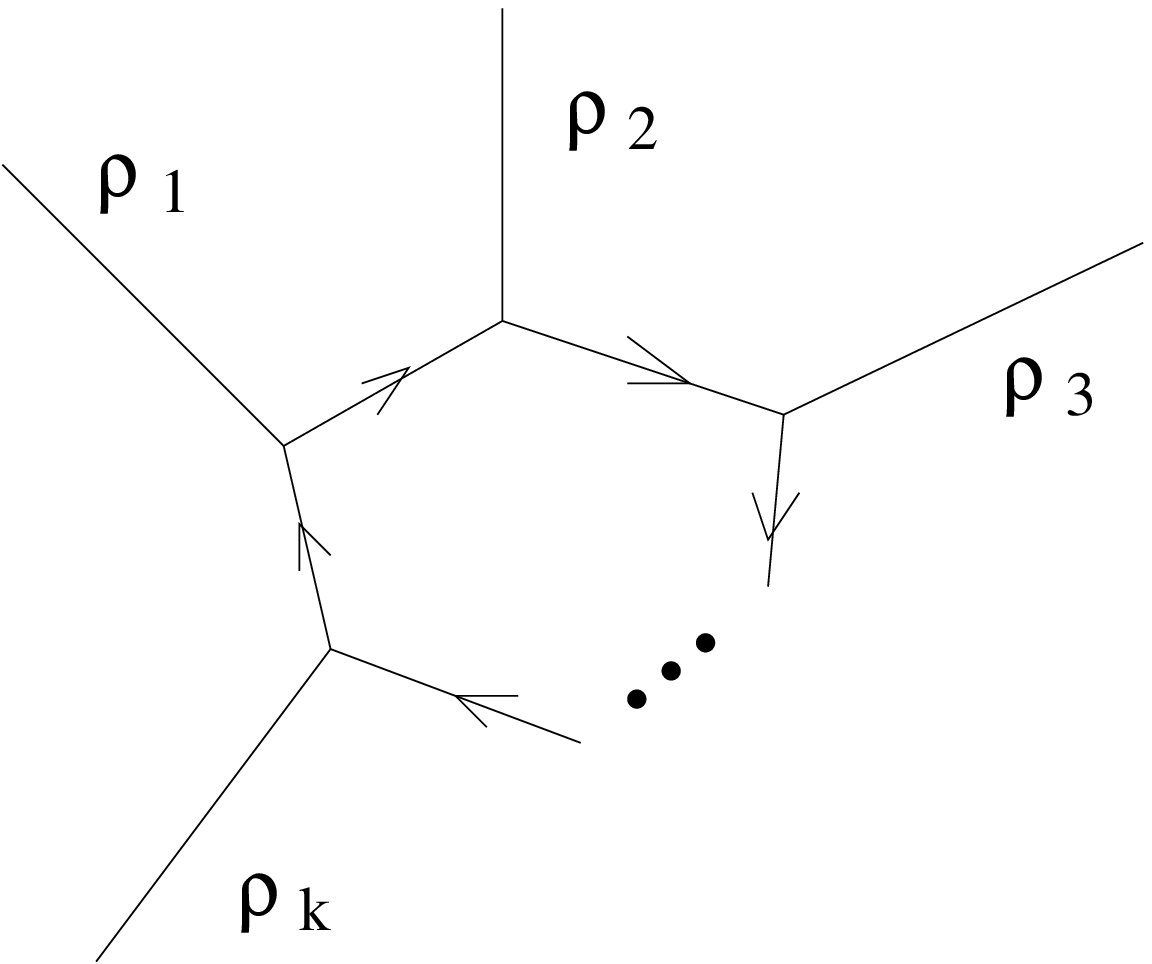,height=1.2in}} 
\ee
Here the representation on the internal loop is the null representation $V^0$. 

We would now like to argue that the quantities \eqref{product-pr} satisfy all the properties of correlation functions of a CFT, so the
prescription \eqref{product-pr} defines a CFT. We recall that CFT can be thought of
as a machine for computing correlation functions. The spectrum of CFT is its primary operator content. In our case
the spectrum is all representations that appear in \eqref{dec'}. A CFT is defined if one has a rule for computing
any $k$-point function. In our case this rule is given by \eqref{product-pr}. There are some consistency requirements
that such a rule must satisfy to be a definition of a $k$-point function of some CFT. These consistency requirements
can be reduced to two: (i) the 4-point function can be decomposed into two 3-point functions, and the representations
that appear in the intermediate channel are all operators in the spectrum; (ii) such decomposition must not
depend on which channel is used; this is also known as the crossing symmetry. 

\subsection{} Let us start by considering the 3-point function given by:
\be\label{3-point}
\lower0.35in\hbox{\epsfig{figure=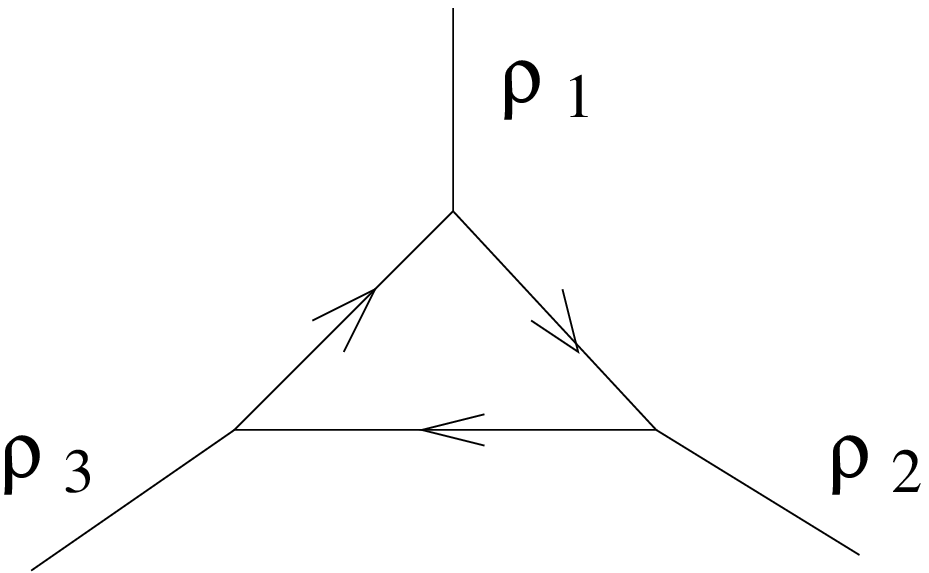,height=0.7in}} 
\ee
Note now that the space of intertwiners from the tensor product of representations $\rho_1,\rho_2,\rho_3$ to the
trivial representation is one dimensional. Indeed, this intertwiner is just the 3-point function of the
corresponding CFT operators, and this is fixed, up to the normalization, by the conformal symmetry. 
Let us fix some normalization of this intertwiner. The normalization is
encoded in the following object:
\be\label{theta}
\theta(\rho_1,\rho_2,\rho_3) = \lower0.4in\hbox{\epsfig{figure=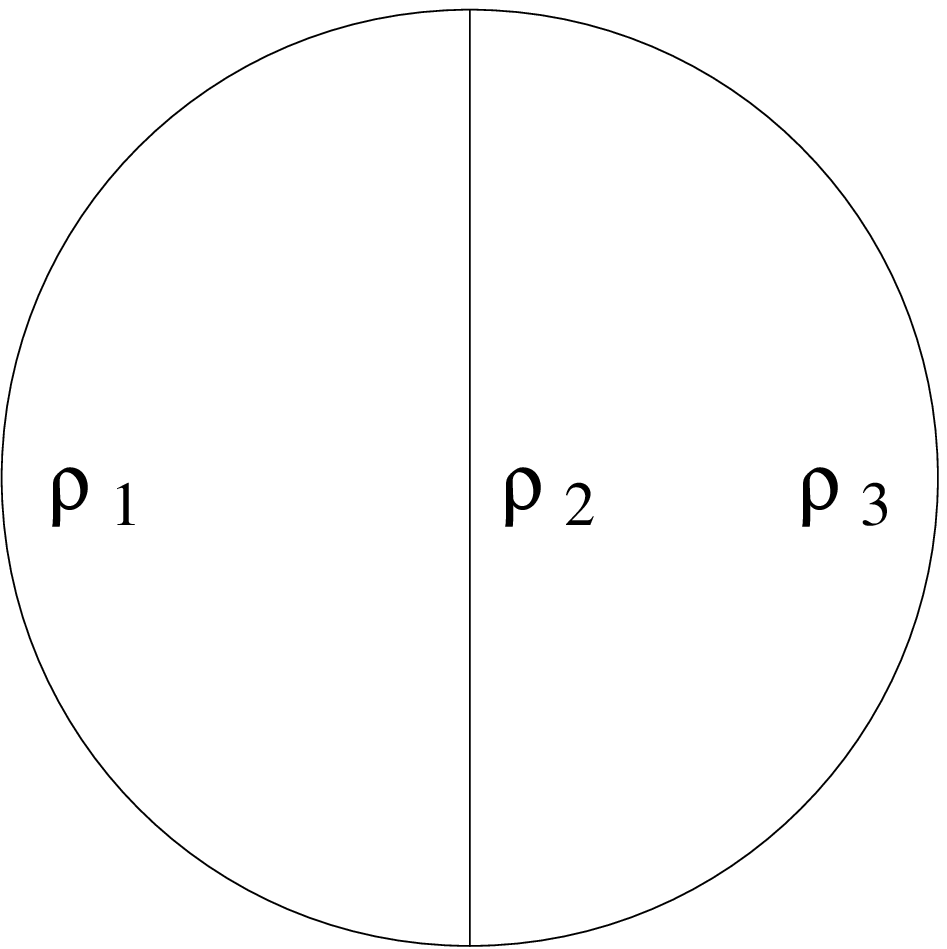,height=0.8in}}
\ee
The object \eqref{3-point} must be proportional to the intertwiner in question:
\be\label{3-point-6j}
\lower0.35in\hbox{\epsfig{figure=3-point.eps,height=0.7in}} \,\, =\,\, \frac{1}{\theta(\rho_1,\rho_2,\rho_3)}\,
\lower0.4in\hbox{\epsfig{figure=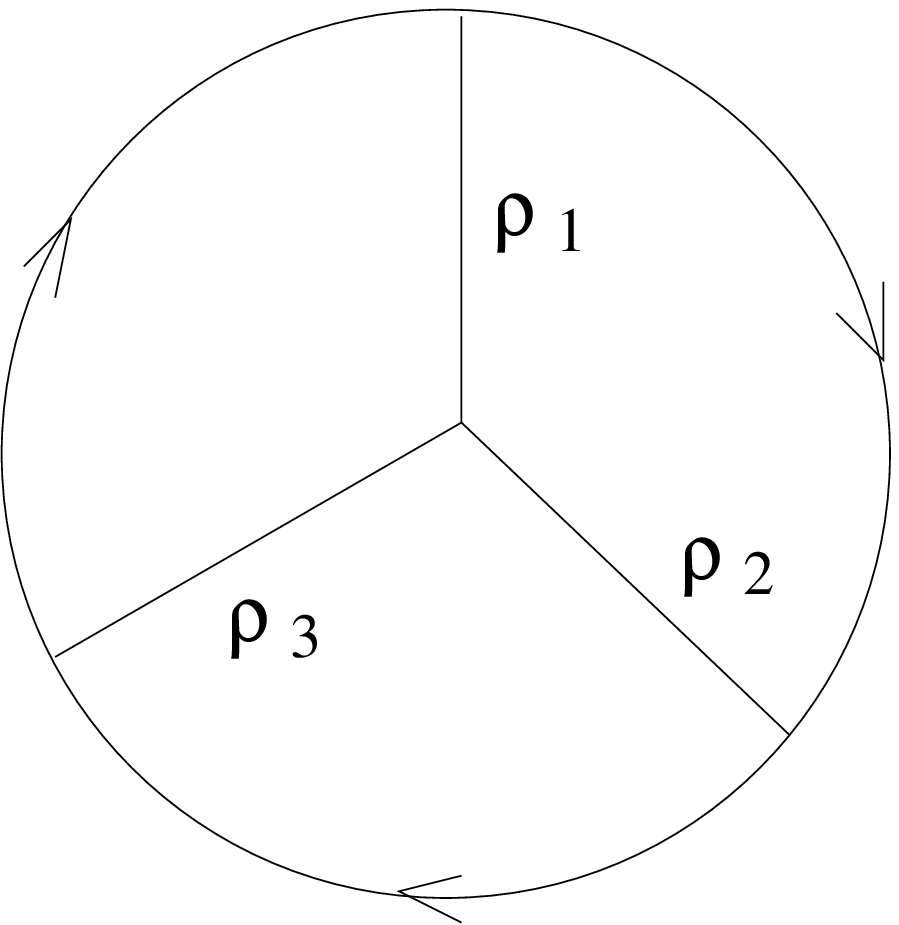,height=0.8in}} \,\, \lower0.4in\hbox{\epsfig{figure=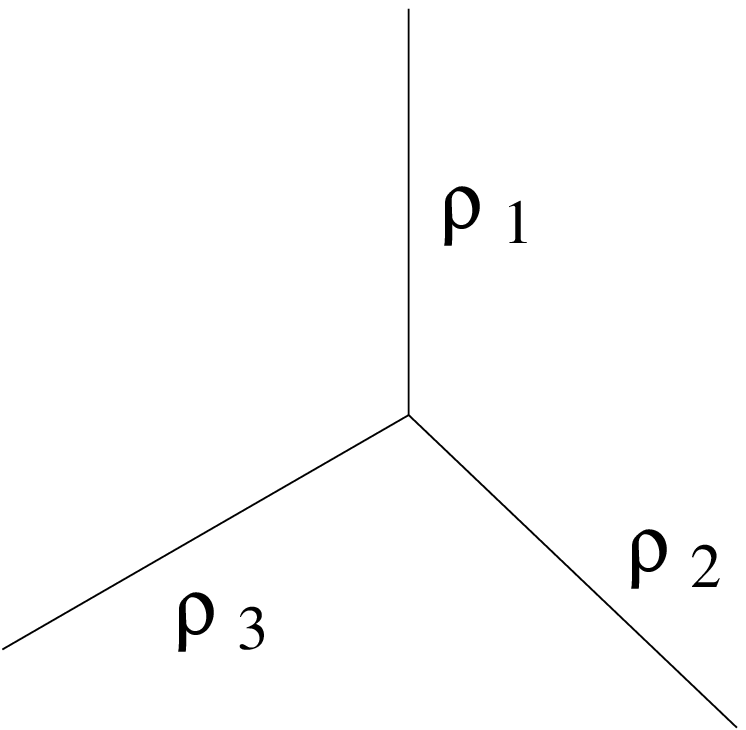,height=0.8in}}
\ee
The coefficient of proportionality can be checked by applying the intertwiner $C^{\rho_1 \rho_2 \rho_3}$ to
both sides of \eqref{3-point-6j}. Thus, we see that the non-trivial part of the 3-point function of our CFT is given by the
6j-symbol. 

\subsection{}
Let us now consider the 4-point function. It is given by the following diagram:
\be\label{4-point}
\lower0.45in\hbox{\epsfig{figure=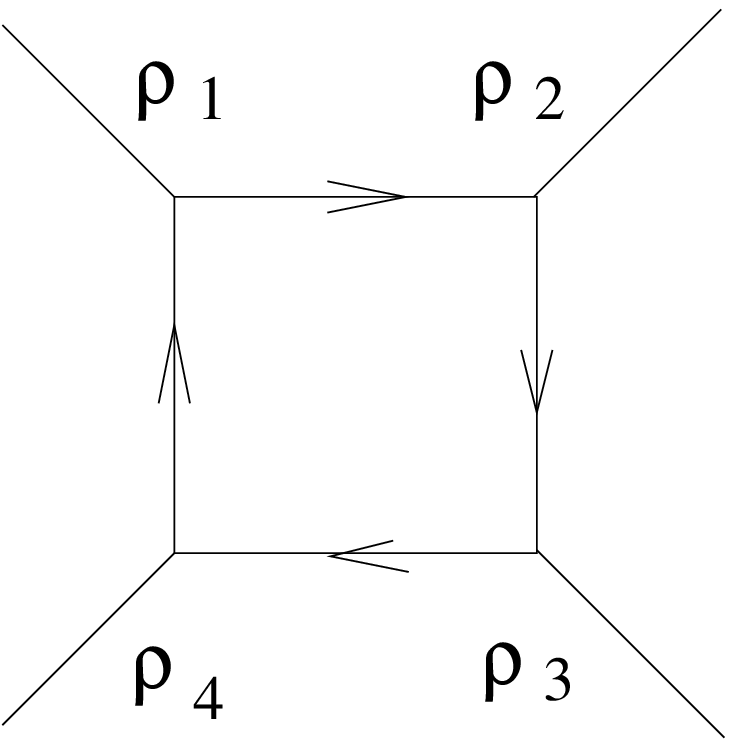,height=0.9in}} 
\ee
Let us now demonstrate that the property (i) described above indeed holds. We start with the following identity:
\be\label{ident}
\lower0.15in\hbox{\epsfig{figure=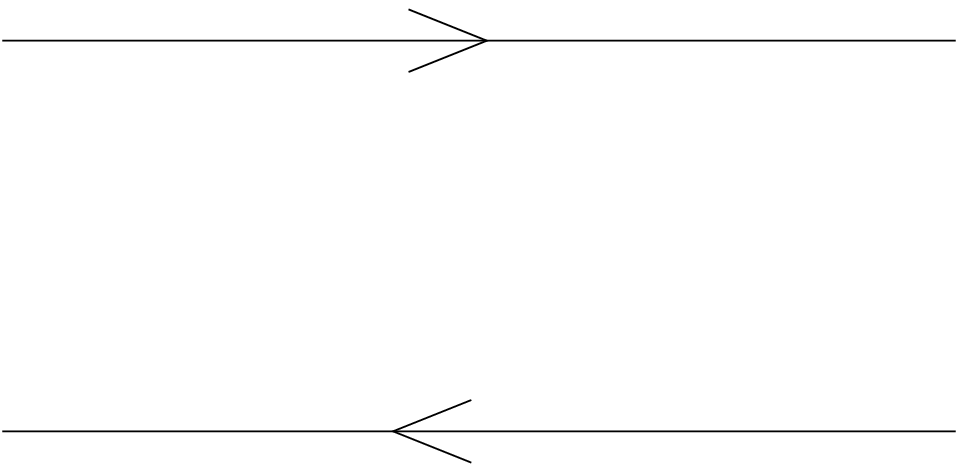,height=0.3in}}\,\, =\sum_\rho  \,\,\, 
\lower0.25in\hbox{\epsfig{figure=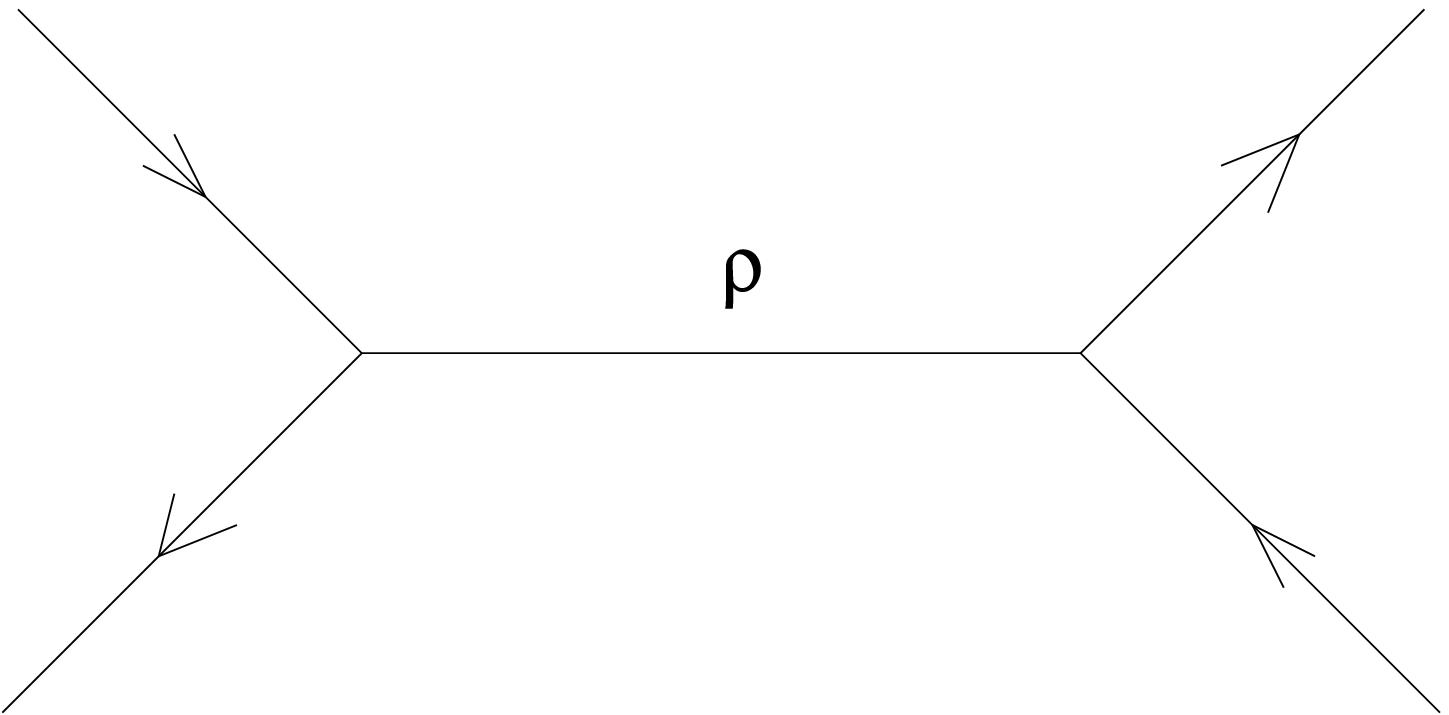,height=0.6in}}
\ee
The quantity on the left hand side is the product of two identity operators: one in $V^0$ one in $\overline{V^0}$, and
the sum on the right hand side is taken over all the representations that appear in the decomposition \eqref{dec'}. 
This identity is proven by attaching the Clebsch-Gordan coefficient to the right of both sides. Using the 
orthogonality \eqref{3j-ort} one gets the Clebsch-Gordan on both sides. Let us now apply the
identity \eqref{ident} to two of the internal lines of the 4-point function \eqref{4-point}. We get:
\be\label{4-point-1}
\lower0.45in\hbox{\epsfig{figure=4-point.eps,height=0.9in}} \,\, = \,\, \sum_\rho \,
\lower0.35in\hbox{\epsfig{figure=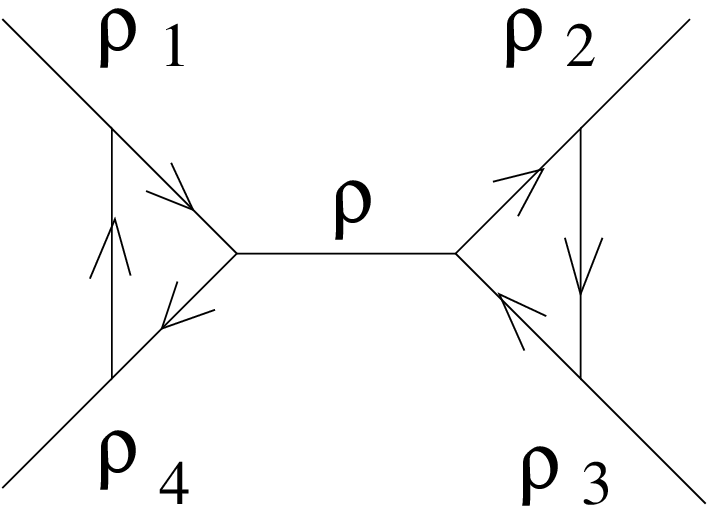,height=0.7in}}
\ee
We thus see that the 4-point function decomposes precisely into two full 3-point functions \eqref{3-point}.
The states appearing as intermediate, that is, the spectrum of the CFT, are all the states appearing
in the decomposition \eqref{dec'}. It is also clear that the same decomposition \eqref{4-point-1} can be applied
in another channel; we thus have the crossing symmetry property. 

To summarize, using only very general representation theoretic identities we have established that the
quantities given by \eqref{product-pr} are correlation functions of some CFT. We did not need 
any explicit form of the Clebsch-Gordan coefficients, or even any details on how the representations
in question are realized. Of course, in order to do any explicit calculations with this formalism such
details are important. The question of interest is, for example, what is the 6j-symbol that
appeared in the expression for the 3-point function \eqref{3-point-6j} and that plays the
role of the structure constant of our CFT. We shall not attempt to calculate this quantity in the
present paper.

\section{Holography}
\label{sec:hol}

We have constructed a CFT from purely representation theoretic considerations. Quantization of the twistor space gave us
a prescription for representing functions from the right hand side of \eqref{dec'} by operators in $V^0$. These
operators $\hat{\mathcal O}_\rho$ are CFT operators, and their correlation functions are computed \eqref{product-pr} 
as the trace of an operator product. The CFT in question ``lives'' on $S^n$, its operators are local operators
$\hat{\mathcal O}_\rho(p), p\in S^n$. A natural question to ask is if there is any theory in $\HH^{n+1}$
that is dual to the CFT constructed. 

As we have seen, the twistor (null) representation $V^0$ can be realized by massless fields in $\HH^{n+1}$. 
We have also seen that all other primaries of our CFT will appear as intermediate states of the
correlation functions of the basic null operator $\hat{\mathcal O}_\rho(p), \rho=0$. For these basic operators we 
shall omit the representation index and denote them simply by $\hat{\mathcal O}(p)$. The correlation 
functions of $\hat{\mathcal O}(p)$ are given by \eqref{product-pr}. Correlation functions of all
other operators can be reconstructed by decomposing those for $\hat{\mathcal O}(p)$ into 
intermediate channels. This suggests that
the dual bulk theory is a theory of a single massless scalar field $\phi$. The situation in the
usual AdS/CFT dualities is analogous: one has only very few bulk fields (SUGRA fields) as compared
to the number of primaries in the boundary CFT. The general boundary correlation function can be
reconstructed via the intermediate channel decomposition.

To address the question what the bulk theory is, let us introduce the generating functional for
the correlation functions of $\hat{\mathcal O}(p)$:
\be\label{gen}
&{}& e^{F[\phi]} = \langle e^{\int_{S^n} dp\, \phi(p) \hat{\mathcal O}(p)} \rangle = \\ \nonumber
&{}& \sum_n \frac{1}{n!} \int_{S^n} dp_1 \ldots dp_n \,\, \phi(p_1) \ldots \phi(p_n) \Tr{\left(\hat{\mathcal O}(p_1) \ldots 
\hat{\mathcal O}(p_n)\right)}.
\ee
We note that, unlike the usual AdS/CFT story, all the correlation functions on the right hand side are, at least
in principle, explicitly calculable from the prescription \eqref{product-pr}. For the usual AdS/CFT dualities
the boundary theory is strongly coupled, so no calculation of the CFT correlators is possible, apart from the
simplest cases protected by the non-renormalization theorems. One obtains the correlators by doing bulk computations.
The situation here is the reverse: we have the boundary theory, and the question is what theory in bulk, if any,
that corresponds to it.

We have argued that the bulk theory must be that of a single massless scalar field. Its action is therefore of the form:
\be\label{action}
S[\phi] = g \int_{\HH^{n+1}} dx \, \left( \phi \Delta_{\HH} \phi + \sum_{k=3}^\infty g_k \phi^k \right).
\ee 
In principle, interaction terms of any order can appear. Assume now that the generating functional \eqref{gen}
has been computed. The corresponding bulk theory $S[\phi]$ is such that, when solving the bulk equations of
motion with boundary value $\phi|_b$ of the field $\phi$ fixed, and substituting this solution into the action,
one gets \eqref{gen}:
\be
S[\phi[\phi|_b]] = F[\phi|_b].
\ee
Once all the correlation functions for $\hat{\mathcal O}(p)$ are known, the bulk action can be computed order by order.
Indeed, one would first fix the coefficient in front of the action \eqref{action} so that the boundary
2-point function one obtains via the holographic prescription is the same way as the CFT one.
In our case $g$ would have to be fixed so that the orthonormality relation \eqref{3j-ort} for the Clebsch-Gordan 
coefficients is reproduced. After the bulk and boundary 2-point functions have been matched, one should compute the 3
-point function using the holographic
prescription. The coupling constant $g_3$ should then be adjusted as to give the correct CFT 3-point function
\eqref{3-point}. One then continues order by order, introducing, if necessary, new interaction terms $g_k \phi^k$ to match
the CFT k-point function. 

The procedure described is probably not very practical, but, at least in principle, possible. It would be of much
more interest to determine the bulk theory some other way. Indeed, the CFT we have constructed is rather simple.
Similar representation theoretic arguments may suggest what the bulk theory is. Unfortunately, we don't have
anything more to say on this issue at present.

\section{Discussion}
\label{sec:disc}

By quantizing the twistor space $\TT$ we have obtained a CFT whose spectrum of primary operators is the spectrum of
representations appearing in the decomposition~\eqref{dec'}. Correlation functions of this CFT are given by
the diagrams \eqref{product-pr}. We have argued that the dual theory is a theory of a single non-interacting massless 
field in AdS. At least in principle, the interactions of the bulk theory can be determined order by order.  

Of course
this is just a toy model of an AdS/CFT duality. Physically interesting dualities involve much more complicated theories.
However, our construction allows for generalizations. An obvious generalization is to take a more complicated group.
For instance, one can take a super-conformal group which would be relevant for holography on AdS cross some internal space.
Let us consider the case of AdS${}_5\times S^5$. The bosonic part of the relevant super-conformal group is $\SO(1,5)\times\SO(6)$.
A generalization of our construction to this case would be to consider as an analog of the space of twistors 
the Grassmanian:
\be
\left(\SO(1,5)/\SO(1,3)\times\SO(2) \right) \times \left(\SO(6)/\SO(4)\times\SO(2)\right).
\ee
When quantized it corresponds to a particular representation of the product group: 
\be\label{V}
V= V^0\otimes \left(\oplus_N V^N \right),
\ee
where $V^N$ are the spherical harmonics representations of $\SO(6)$. When realized in the space of functions on $\HH^5$
these representations are those of massive fields. The spectrum of masses is given by the quadratic Casimir of $\SO(6)$:
$m^2=N(N+4)$. These are some of the modes of the relevant supergravity theory on AdS${}_5$. The operator content
of the corresponding CFT can be found by taking the tensor product $V\otimes\overline{V}$ and decomposing it into irreducible
representations. The bulk theory would in this case contain all the massive KK modes described by $V$ \eqref{V}. 
Thus, by taking a more complicated group, and applying our construction, one obtains a whole zoo of (rather simple) bulk/boundary
dualities. Of course, as it stands, to find the bulk theory one needs to perform a laborious order by order computation.

Another possible generalization would be to introduce a parameter to the theory. The CFT constructed should be
thought of as an analog of the ${\mathcal N}=4$ super Yang-Mills at infinite coupling and infinite $N$. 
In this case the usual SUGRA description is applicable; similarly, there is no parameter to be tuned. 
The CFT of the present paper was obtained using the non-commutative geometry type construction. One
might foresee a more complicated parameter-dependent quantization of the twistor space. For instance,
one could consider a sigma model on the twistor space. Open sigma models are known to be
responsible for non-commutativity, thus making them a good candidate for a possible generalization
of the present story. It is natural to speculate that after a parameter has been introduced, the dual bulk theory would 
contain higher derivative corrections, similar to the $\alpha'$ corrections to the SUGRA action.

An important point left out by our discussion is if the dual theory is gravity. In the construction presented it is
certainly not: the bulk theory is a theory of a single massless field. Indeed, we have quantized the twistor space
corresponding to a fixed conformal structure of $S^4$. As the experience with AdS/CFT correspondence suggests, the 
graviton in bulk should be dual to the stress-energy tensor of the boundary theory. Such an object can only arise
if one allows variations of the conformal structure on $S^4$. Thus, to incorporate gravity one would need to
quantize not simply the twistor space, but twistors together with conformal structures. This is clearly 
a much more complicated problem. We note that non-linear graviton of Penrose, see~\cite{book}, may be of relevance 
here. It would be of interest to consider these issues. Similarly, one may attempt to quantize not just the
twistor space, but the space of holomorphic vector bundles on $\TT$. This can potentially lead to
Yang-Mills-like theories. Thus, we would like to express hope that more realistic CFT's can be built using a
construction similar to the one presented. 

To conclude, the story presented here is certainly not complete. However, the construction of the present paper is simple 
and allows for generalizations in various directions. It seems worth to be developed further.
Probably the most important problem is to understand the principle that fixes the bulk theory. The story presented here 
would become much more exciting if there is a way to write the bulk theory without recourse to the boundary CFT
computations. We would like to speculate that the Penrose transform \cite{Gindikin} is of crucial 
importance here. When the bulk theory is better understood, it would be of great interest to apply the
construction to more realistic groups/theories. We hope to return to these questions in future works.

\subsection*{Acknowledgments} I would like to thank R. Penrose for illuminating correspondence and G. Arutyunov for discussions.
Thanks to L. Freidel and G. Arutyunov for reading the manuscript and for their extremely helpful comments.

\end{document}